\documentclass[pre,aps,eqsecnum,twocolumn,showpacs]{revtex4}
\usepackage[centertags]{amsmath}
\usepackage{amssymb}
\usepackage{amsthm}
\begin{document}

\title{Stochastic Energetics of Quantum Transport}
\author{Pulak Kumar Ghosh and Deb Shankar Ray{\footnote{Email Address:
pcdsr@mahendra.iacs.res.in}}} \affiliation{Indian Association for
the Cultivation of Science, Jadavpur, Kolkata 700 032, India}

\begin{abstract}
We examine the stochastic energetics of directed quantum transport
due to rectification of non-equilibrium thermal fluctuations. We
calculate the quantum efficiency of a ratchet device both in
presence and absence of an external load to characterize two
quantifiers of efficiency.  It has been shown that the quantum
current as well as efficiency in absence of load (Stokes efficiency)
is higher as compared to classical current and efficiency,
respectively, at low temperature. The conventional efficiency of the
device in presence of load on the other hand is higher for a
classical system in contrast to its classical counterpart. The
maximum conventional efficiency being independent of the nature of
the bath and the potential remains the same for classical and
quantum systems.
\end{abstract}
 \maketitle

\section{Introduction}
Forced thermal ratchet has been the paradigm for rectification of
non-equilibrium fluctuations\cite{ajd,mag1,jul,mag} for usable work.
In the simplest possible terms it represents a Brownian particle in
a periodic potential under overdamped condition  which exhibits a
net drift provided the system is subjected to an external force with
zero mean and with sufficient correlation so that the detailed
balance is lost and the symmetry of the device is broken. Over the
years considerable attention has been devoted to this area to
understand functioning of molecular motors active in muscle
contraction \cite{ser,val,how,tso}, useful separation of particle
\cite{rou} theoretical issues involving second law
\cite{van,lef,sek,kam} and many other aspects\cite{gla,stu}. Since a
ratchet device is a typical machine which works at a mesoscopic
level in converting heat drawn from non-equilibrium fluctuations
into work, attempts have been made to quantify the efficiency of
such a machine. For example, Sekimoto \cite{sek} has proposed a
method for studying several variants of thermal ratchet model
analyzed also by others\cite{kam1}. Magnasco \cite{mag} has
considered a Sziland's heat engine and suggested an expression for
net power consumed by such an engine. The work of Julicher et al
\cite{jul} had provided the estimate of total energy consumption. An
interesting generalization of definition of efficiency had been
proposed by Der\'{e}nyi \textit{et al} \cite{der} for motors without
load.

We address in this paper the problem of stochastic energetics of a
forced thermal ratchet in a quantum mechanical context. A Brownian
particle being a microscopic object, the quantum effect is likely to
be significant in appropriate situation, e. g, in the transport of
quantum particles in quantum wires, superionic\cite{bru,gei,dic}
conductors and in other nanodevices\cite{lin}, particularly at low
temperature and other important issues\cite{doe,luc,mie}. To this
end a number of attempts on quantum ratchet device have been made.
Reimann \textit{et al} \cite{rei1} investigated adiabatically rocked
ratchet system to show that quantum corrections enhance classical
transport at low temperature. Two models of quantum ratchet have
also been proposed by Yukawa \textit{et al} \cite{yuka}. Based on
the perturbative approach Scheidl and Vinokur \cite{sche} have
investigated quantum Brownian motor in ratchet potentials to
identify the characteristic scales of response functions of the
system. Carlo \textit{et al} \cite{carlo} have studied a typical
model quantum chaotic dissipative ratchet to analyze the directed
transport from a quantum strange attractor. Keeping in view of this
development we note that although quantum ratchet device has been
the object of interest for some time, its efficiency (i. e., quantum
efficiency) has largely remained unexplored. Based on the quantum
Langevin equation which implies an interplay between several forces
we analyze here the energetics of directed transport by taking into
consideration of how the transducer which characterizes the state of
system mediates the energy among the basic components of a forced
ratchet, i. e., the external system, the load, and the heat bath. It
also important that frictional dissipative energy in course of
directed motion must also be counted as a part of expenditure of
useful energy for rectification of Brownian motion. This implies
that one can also envisage a kind of Stokes efficiency in absence of
load. We take into account of these considerations in our
exploration of quantum energetics in presence and absence of load to
characterize two distinct quantifiers of efficiency in a quantum
ratchet device. A relevant pertinent point that needs attention in
this context is that although quantization, in principle, adds new
elements into the theory, it is important that quantization must not
break the symmetry of the device, i. e., it should not create a new
load or tilt on the potential or break the detailed balance.
Secondly, forcing must be unbiased so that after appropriate
averaging over time or ensemble, no directional component should
appear as a fictitious drift. With these considerations for
thermodynamic consistency we analyze the efficiency and current
generation in a quantum ratchet in relation to total consumption of
energy and dissipation both in presence and absence of external
load.

\section{A quantum dynamics in a spatially periodic potential at equilibrium}

We consider a particle of mass $m$ moving in a periodic classical
potential $V(x)$. The particle is coupled to a set of harmonic
oscillators of unit mass acting as a bath. This is represented by
the following system-reservoir Hamiltonian \cite{zwa,mori}
\begin{equation}\label{2.1}
\hat{H}=\frac{\hat{p}^2}{2 m}+V(\hat{x})+\sum_{j=1}^N \left\{
\frac{\hat{p}^2_j}{2}+\frac{1}{2} \kappa_j (\hat{q}_j-\hat{x})^2
\right\}
\end{equation}
Here $\hat{x}$ and $\hat{p}$ are the coordinate and momentum
operators of the particle and $\{\hat{q}_j, \hat{p}_j\}$ are the set
of coordinate and momentum operators for the reservoir oscillators
coupled linearly through the coupling constants ${\kappa}_j
(j=1,2,...)$. For the spatially periodic potential, we have
$V(x)=V(x+L)$,where $L$ is the length of the period.The coordinate
and momentum operators follow the usual commutation rules
$\{\hat{x}, \hat{p}\}=i\hbar$ and $\{\hat{q}_i,
\hat{p}_j\}=i\hbar{\delta}_{ij}$. Eliminating the bath degrees of
freedom in the usual way we obtain the operator Langevin equation
for the particle
\begin{equation}\label{2.2}
m \ddot{\hat{x}}+\int^{ t}_0
d{t'}\gamma({t}-{t'})\dot{\hat{x}}({t'})+V'(\hat{x}) =
\hat{\Gamma}({t})
\end{equation}
(Overdots refers to differentiation with respect to time ${t}$)
where noise operator $\hat{\Gamma}({t})$ and the memory kernel are
given by
\begin{equation}\label{2.3}
\hat{\Gamma}({t}) =
\sum_j\left[\{\hat{q}_j(0)-\hat{x}(0)\}\kappa_j\cos\omega_j{t}
+\kappa_j^{1/2}\hat{p}_j(0) \sin\omega_j{t}\right]
\end{equation}
and
\begin{equation}\label{2.4}
\gamma({t}) =\sum_j\kappa_j\cos\omega_j{t}
\end{equation}
respectively, with $\kappa_j=\omega_j^2$

Following Ref. \cite{db1} we then carry out a quantum mechanical
average $\langle...\rangle$ over the product separable bath modes
with coherent states and the system mode with an arbitrary state at
${t} =0$ in Eq.(\ref{2.2}) to obtain a generalized quantum Langevin
equation\cite{db4,d41,db1} as
\begin{equation}\label{2.5}
m \ddot{{x}} + \int_0^{{t}}d{t'}\gamma({t}-{t'})
\dot{{x}}({t'})+{V'}({x})= {\Gamma}({t}) + {Q}({x}
,{\langle\delta\hat{x}^n\rangle})
\end{equation}
where the quantum mechanical mean value of the position operator
$\langle\hat{x}\rangle ={x}$ and
\begin{equation}\label{2.6}
{Q}({x} ,{\langle\delta\hat{x}^n\rangle)} = {V'}({x}) - {\langle
V'(\hat{x})\rangle}
\end{equation}
which by expressing  $ \hat{x}({t}) = {x}({t}) + \delta\hat{x}({t})$
in $V(\hat{x})$ and using a Taylor series expansion around ${x}$ may
be rewritten as
\begin{equation}\label{2.7}
{Q}({x} ,{\langle\delta\hat{x}^n\rangle}) =-\sum_{n\geq
2}\frac{1}{n!}{V}^{n+1}({x}){\langle\delta\hat{x}^n\rangle}
\end{equation}
The above expansion implies that the nonzero anharmonic terms beyond
$n\geq 2$ contain quantum dispersions
${\langle\delta\hat{x}^n\rangle}$. Although we develop this section
in general terms, we are specifically concerned here typically with
periodic nonlinear potentials of the type $\sin\frac{2\pi x}{L}$ or
$\cos\frac{2\pi x}{L}$ or their linear combinations and the like
which have been used earlier in several contexts. The nonlinearity
of the potential is an important source of quantum correction in
addition to the quantum noise of the heat bath. The calculation of
${Q}$ rests on the quantum correction terms
${\langle{\delta\hat{x}^n}\rangle}$ which one determines by solving
a set of quantum correction equations as given in the next section.
Furthermore the c-number Langevin\cite{d41,db1,db4} force is given
by
\begin{equation}\label{2.8}
{\Gamma}({t}) =
\sum_j\left[\langle\hat{q}_j(0)\rangle-\langle\hat{x}(0)\rangle
\kappa_j\cos\omega_j{t} +\kappa_j^{1/2}\hat{p}_j(0)
\sin\omega_j{t}\right]
\end{equation}
which must satisfy noise characteristics of the bath at equilibrium,
\begin{eqnarray}
\langle {\Gamma}({t}) \rangle_S & = &
0\label{2.9}\\
\langle {\Gamma}({t}) {\Gamma}({t'}) \rangle_S &=& \frac{1}{2}
\sum_j \kappa_j\; \hbar \omega_j \left( \coth \frac{\hbar
\omega_j}{2 k T} \right) \cos \omega_j
({t}-{t'})\nonumber
\\\label{2.10}
\end{eqnarray}
Eq.(\ref{2.10}) expresses the quantum fluctuation-dissipation
relation. The above conditions(2.9-2.10) can be fulfilled provided
the initial shifted co-ordinates
$\{\langle\hat{q}_j(0)\rangle-\langle\hat{x}(0)\rangle\}$ and
momenta $\langle{\hat{p}_j}(0)\rangle$ of the bath oscillators are
distributed according to the canonical thermal Wigner distribution
\cite{wig,hil} of the form
\begin{eqnarray}\label{2.11}
&&P_j([\langle\hat{q}_j(0)\rangle-\langle\hat{x}(0)\rangle],
\langle\hat{p}_j(0)\rangle) \nonumber \\
 &=& N
\exp\left\{-\;\frac{\frac {1}{2}\langle\hat{p}_j(0)\rangle^2 + \frac
{1}{2}\kappa_j[\langle\hat{q}_j(0)\rangle-\langle\hat{x}(0)\rangle]^2}
{\hbar\omega_j[{n}(\omega_j) + \frac{1}{2}]}\right\}\nonumber\\
\label{2.11}
\end{eqnarray}
 so that the statistical averages $\langle...\rangle_s
$ over the quantum mechanical mean value $O$ of the bath variables
are defined as
\begin{equation}\label{2.12}
\langle O_j \rangle_s=\int O_j\; P_j\; d\langle
\hat{p}_j(0)\rangle\;
d\{\langle\hat{q}_j(0)\rangle-\langle\hat{x}(0)\rangle\}
\end{equation}
Here ${n}(\omega)$ is given by Bose-Einstein distributions
$(e^{\frac{\hbar\omega}{kT}}-1)^{-1}$. $P_j$ is the exact solution
of Wigner equation for harmonic oscillator \cite{wig,hil} and forms
the basis for description of the quantum noise characteristics of
the bath kept in thermal equilibrium at temperature $T$. In the
continuum limit the fluctuation-dissipation relation (\ref{2.10})
can be written as
\begin{eqnarray}\label{2.13}
&&\langle{\Gamma}({t}){\Gamma}({t'})\rangle \nonumber
\\
&=&\frac{1}{2}\;\int_0^\infty d\omega \;\kappa(\omega)\;\rho(\omega)
\;\hbar\omega\;
\coth({\frac{\hbar\omega}{2kT}})\;\cos{\omega({t}-{t'}})\nonumber
\\\label{2.13}
\end{eqnarray}
where we have introduced the density of the modes $\rho(\omega)$.
Since we are interested in the Markovian limit in the present
context, we assume  $\kappa(\omega)\rho(\omega)
=\frac{2}{\pi}\gamma$, Eq.(\ref{2.13}) then yields

\begin{equation}\label{2.14}
\langle{\Gamma}({t}){\Gamma}({t'})\rangle =2 {D}_q\delta({t}-{t'})
\end{equation}

with

\begin{equation}\label{2.15}
{D}_q
=\frac{1}{2}\gamma\hbar\omega_0\coth{\frac{\hbar\omega_0}{2kT}}
\end{equation}
(The passage from Eq.(\ref{2.13}) to Eq.(\ref{2.14}) is given in the
appendix A.)

 $\omega_0$ refers to static frequency limit.
 Furthermore from Eq.(\ref{2.4}) in the continuum limit we have

\begin{equation}\label{2.16}
\gamma({t}-{t'}) = \gamma\;\delta({t}- {t'})
\end{equation}

$\gamma$ is the dissipation constant in the Markovian limit. In
this limit Eq.(\ref{2.5}) therefore reduces to

\begin{equation}\label{2.17}
m \ddot{{x}} + \gamma \dot{{x}}+{V'}({x})= {\Gamma}({t }) + {Q}({x}
,{\langle\delta\hat{x}^n\rangle})
\end{equation}

It is useful to work with dimensionless variables for the present
problem to keep track of the relations between the scales of
energy, length and time. The period $L$ of the periodic potential
$V(x)$ determines in a natural way the characteristic length scale
of the system. Therefore the position of the Brownian particle is
scaled as

\begin{equation}\nonumber
\overline x={x}/L
\end{equation}

Next we consider the timescales of the system. In absence of the
potential and the noise term the velocity of the particle
$\dot{x}({t})\sim \exp(-{t}/\tau_L)$ with $\tau_L=m/\gamma$, which
represents the correlation time scale of the velocity the Brownian
particle. To identify the next characteristic time $\tau_0$ we
consider the deterministic overdamped motion due to the potential as
$\gamma\frac{d{x}}{d{t}} = -\frac{d{V}({x})}{d{x}}$. Then $\tau_0$
is determined from $\gamma\frac{L}{\tau_0}=-\frac{{\Delta V}\ }{L}$
as $\tau_0=\frac{\gamma L^2}{{\Delta V}}$ where ${\Delta V} $ is the
barrier height of the original potential. Hence time is scaled as
$\overline t = \frac{{t}}{\tau_0}$. Furthermore the potential, the
noise and the quantum correction terms are re-scaled as
$\overline{V}(\overline x)={V}({x})/\Delta{V}$, $ \overline
{\Gamma}(\overline t)={\Gamma}(t)/(\Delta{V}/L)$ and
${Q}/(\Delta{V}/L)$, respectively.

Hence dimensionless quantum Langevin equation reads as

\begin{equation}\label{2.18}
\mu^*\ddot{\overline x}+\dot{\overline x}=\overline{f}(\overline
x)+\overline{\Gamma}(\overline t)
\end{equation}

Here over-dot(.) refers to differentiation with respect to scaled
time $t$. Dimensionless mass
$\mu^*=\frac{m}{\gamma\tau_0}=\frac{\tau_L}{\tau_0}$ and

\begin{equation}\label{2.19}
\overline{f}(\overline x)=-\overline{V}'(\overline x)+
\overline{Q}(\overline x ,\overline{\langle\delta\hat{x}^n\rangle})
\end{equation}

The noise properties of the quantum bath are then rewritten as

\begin{equation}\nonumber
\langle\overline{\Gamma(\overline t)}\rangle_s = 0
\end{equation}
\begin{equation}\nonumber
\langle\overline{\Gamma(\overline t)}\overline{\Gamma(\overline
t')}\rangle_s =2 \overline{D}_q\delta(\overline t-\overline{t}')
\end{equation}
where

\begin{equation}\nonumber
\overline{D}_q=\frac{\frac{1}{2}\hbar\omega_0\coth{\frac{\hbar\omega_0}{2kT}}}{{\Delta
V }}
\end{equation}
From now onwards we drop the over-bars from all parameters and
variables for simplicity.
 It may be shown\cite{db4,d1} that quantum stochastic dynamics Eq.(\ref{2.18})
 does not generate drift motion to a preferential direction.
Since the quantum correction $Q(x,\langle\delta\hat{x}^n\rangle)$ ,
as expected, can not break the detailed balance in the quantum
system, nor the symmetry of the potential. This conclusion is an
important check of the present formalism for a correct description
of the equilibrium and thermodynamic consistency.

\section{Quantum transport induced by zero mean external fluctuation}

\subsection{General features}
Since equilibrium thermal fluctuations due to heat bath can not
break detailed balance in the quantum stochastic dynamics, we
introduce an external derive with zero mean and with sufficient
correlation to generate drift motion on average in one direction. To
analyze the energetics of directed quantum transport, we now
introduce an external load to work against the global motion of the
forced thermal ratchet system. From Eq.(\ref{2.18}) it follows that
the dynamics of the particle under overdamped condition is described
by the scaled equation (we have dropped the over-bar)

\begin{equation}\label{3.1}
\dot{x}=f(x)+\Gamma(t)+A(t)-\frac{\partial V_l}{\partial x}
\end{equation}

The quantum mechanical mean of the position operator, $x$ represents
the state of the energy transducer, that is the state of the
ratchet. $\Gamma(t)$ is the internal quantum noise of the thermal
bath with the properties as noted earlier. $A(t)$ is an external
field with temporal period $\tau$,  $A(t+\tau)=A(t)$, in the present
problem. We consider $A(t)=A_0 \sin{wt}$. It is important to note
that for a movement of transducer in a preferential direction $A_0$
must lie between two threshold values, $max_x f(x)$ and -$min_x
f(x)$ \cite{mag}. $\frac{\partial V_l}{\partial x}=l$, is a load
against which transducer performs work. The quantum nature of the
problem therefore manifests itself in two ways; first, through
quantum corrections in $f(x)$ which we consider, in principle, to
all orders and secondly in quantum diffusion coefficient $D_q$ for
the noise of the bath.

The Fokker-Planck  equation corresponding to Eq.(\ref {3.1}) is
given by

\begin{equation}\label{3.2}
 \frac{\partial P(x,t)}{\partial
t}=-\frac{\partial J(x,t)}{\partial x}
\end{equation}
where
\begin{equation}\label{3.3}
J(x,t)=-D_q \frac{\partial P(x,t)}{\partial x}+[f(x)+A(t)-l]P(x,t)
\end{equation}
If forcing frequency is very low, there is enough time for the
system to reach the steady state during the period $\tau$ and the
above equation can be solved analytically for $J$ as a function of
$A$, using period boundary and normalization conditions
\begin{equation}\label{3.4}
 P(x+1)=P(x)  \;\;\; ;\;\;\; \int_c^{c+1}P(x)dx=1\;\;\; ;\;\;\;
\end{equation}
We then obtain
\begin{eqnarray}\label{3.5}
 &&J(A)\nonumber\\
 &=&\frac{\exp{[\psi(1)]}-\exp{[\psi(0)]}}{N\left[
 \{\exp{[\psi(1)]}\}[\int_0^1 \exp{[\psi(x)]}dx
 -C_2]+C_2\exp{[\psi(0)]}\right]}\nonumber\\
 \label{3.5}
\end{eqnarray}
where
\begin{equation}\label{3.6}
N=\frac{1}{D_q}\int_0^1 \exp{[\psi(x)]}dx
\end{equation}
\begin{equation}\label{3.7}
C_2=\frac{\int_0^1 \exp{[\psi(x)]}dx\int_0^x dy
\exp{[\psi(y)]}}{\int_0^1 \exp{[\psi(x)]}dx}
\end{equation}
\begin{equation}\label{3.8}
\psi(x)=\int_c^x \frac{f(y)+A-l}{D_q}dy; \;\;\; \psi(L)=\int_c^1
\frac{f(x)+A-l}{D_q}dx
\end{equation}
The average current over a forcing period is given by
\begin{equation}\label{3.9}
J_{av}=\frac{1}{\tau}\int_0^{\tau} J(A(t))dt
\end{equation}
Average square wave current of amplitude $A_0$ is given by
\begin{equation}\label{3.10}
J_{sqr}=\frac{1}{2}[J(A_0)+J(-A_0)]
\end{equation}

 We now proceed to analyze the current
under non-equilibrium condition and the related quantum effects. One
of the prime quantities for this analysis is the potential $V(x)$ or
the corresponding force term $f(x)$ given by
\begin{eqnarray}\label{3.11}
f(x)&=&-[V'(x)- Q(x ,\langle\delta\hat{x}^n\rangle)] \nonumber
\\
&=&-\frac{\partial}{\partial x}[V(x)+\sum_{n\geq
2}\frac{1}{n!}V^n(x) \langle \delta\hat{x}^n\rangle]
\end{eqnarray}
The quantum correction terms can be determined as follows. We return
to the operator equation (\ref{2.2}) and put $\hat{x}( t)={x}( t)+
\delta\hat{x}( t)$ and $\hat{p}( t)={p}( t)+ \delta\hat{p}( t)$
where ${x}( t)=\langle\hat{x}({t})\rangle$ and ${p}(
t)=\langle\hat{p}({t})\rangle$ are the quantum mechanical mean
values of the operators $\hat{x}$ and $\hat{p}$ respectively. By
construction $[\delta\hat{x},\delta\hat{p}]=i\hbar$ and
$\langle\delta \hat{x}\rangle=\langle\delta \hat{p}\rangle=0$. We
then obtain the quantum correction equation
\begin{eqnarray}\label{3.12}
&&m \delta\ddot{\hat{x}}+\int^{{t}}_0
d{t}'\gamma({t}-{t}')\delta\dot{\hat{x}}({t'})
+{V}''({x})\delta\hat{x}\nonumber
\\
&+& \sum_{n\geq 2}\frac{1}{n!}{V}^{n+1}( x)(\delta\hat{x}^n-
{\langle \delta\hat{x}^n\rangle})=
\hat{\Gamma}({t})-{\Gamma}({t})\nonumber
\\\label{3.12}
\end{eqnarray}
Again in the overdamped limit we discard  the inertial term
$m\delta \ddot{\hat{x}}$. We then perform a quantum mechanical
average with initial product separable coherent states of the
oscillators of the bath only to get rid of the internal noise term
and to obtain the reduced operator equation for the system as

\begin{equation}\label{3.13}
\gamma\delta\dot{\hat{x}} +{V}''({ x})\delta\hat{x}+ \sum_{n\geq
2}\frac{1}{n!}{V}^{n+1}( x)(\delta\hat{x}^n- \langle
{\delta\hat{x}^n\rangle})=0
\end{equation}

With the help of (\ref{3.13}) we then obtain the equations for
${\langle\delta\hat{x}^n(t)\rangle}$

\begin{equation}\label{3.14}
\frac{d}{d
t}{\langle\delta\hat{x}^2\rangle}=\frac{1}{\gamma}\left[-2 V''( x)
{\langle\delta\hat{x}^2\rangle} - V'''(
x){\langle\delta\hat{x}^3\rangle}\right]
\end{equation}
\begin{eqnarray}\label{3.15}
\frac{d}{dt}{\langle\delta\hat{x}^3\rangle}&=&\frac{1}{\gamma}\left[-3
V''( x){\langle\delta\hat{x}^3\rangle} -\frac{3}{2} V'''(
x){\langle\delta\hat{x}^4\rangle}\right]\nonumber \\
&+&\frac{3}{2\gamma} V'''(x){\langle\delta\hat{x}^2\rangle}^2
\end{eqnarray}
and so on. Taking into account of the leading order contribution
${\langle\delta\hat{x}^2\rangle}$ explicitly we may write (it is
easy to observe that each successive order of quantum correction
decreases by a factor of $O(1/\gamma)$ which implies that a leading
order contribution is sufficient in the overdamped limit)
\begin{equation}\label{3.16}
d{\langle\delta\hat{x}^2\rangle}=-\frac{2}{\gamma} V''(
x){\langle\delta\hat{x}^2\rangle} d t
\end{equation}
The overdamped deterministic motion gives $\gamma d x=- V'( x)dt$
which when used in (\ref{3.16}) yields after integration
\begin{equation}\label{3.17}
{\langle\delta\hat{x}^2\rangle}=\Delta_q[ V'( x)]^2
\end{equation}
where $\Delta_q=\frac{{\langle\delta\hat{x}^2\rangle}_{x_c}}{[ V(
x_c)]^2}$ and $ x_c$ is a quantum mechanical mean position at which
${\langle\delta\hat{x}^2\rangle}$ become minimum, i. e.,
${\langle\delta\hat{x}^2\rangle}_{x_c}=\frac{1}{2}\hbar/\omega_0$,
$\omega_0$ being defined in Eq.( \ref{2.15})

In present problem we consider an asymmetric potential of period
$2\pi$,
\begin{equation}\label{3.18}
{V}( x)=-\sin{ x}-0.25\sin{2 x}
\end{equation}
The reference point $x_c$ can be determined by setting
$\frac{d{\langle\delta\hat{x}^2\rangle}}{d x}=0$ and quantum
correction up to the leading order and the potential force are given
by
\begin{equation}\label{3.19}
{Q}( x ,{\langle\delta\hat{x}^n\rangle})=-\Delta_q
[{V}'({x})]^2[{V}'''( x)]
\end{equation}
and
\begin{equation}\label{3.20}
 f( x)=-[ V'( x)+\Delta_q
 V'''( x)[ V'( x)]^2]
\end{equation}
respectively, where $\Delta_q=\frac{2\hbar}{\omega_0}$. We now
emphasize an important point. If the potential is symmetric, then
the quantum correction in Eq.(\ref{3.20}) is an odd function just as
${V}'( x)$. This implies that quantum correction to classical
potential has not destroyed the inversion symmetry of ${V}( x)$.
Thus the approximation in deriving the leading order quantum effect
is consistent with symmetry requirement of the problem. It also
clear that if potential is periodic then the contribution due to
quantum correction to the classical potential i., e., $\int_0^ x Q(x
,{\langle\delta\hat{x}^n\rangle})dx$ is a periodic function of $ x$.
Assuming the form of potential of Eq.(\ref{3.18}), the expression
for quantum correction after properly scaling (as described in the
Sec.II) is given by
\begin{eqnarray}
Q(x ,\langle\delta\hat{x}^n\rangle)&=&-\Delta_q[\cos^3{2\pi
x}+0.5\cos^3{4\pi x}\nonumber
\\
 &+&3\cos^2{2\pi x}\cos{4\pi x}
+2.25\cos{2\pi x}\cos^2{4\pi x}]\nonumber\\
\label{3.21}
\end{eqnarray}
Quantum correction of the potential is entirely due to nonlinearity.
Physically the correction terms account for the quantum fluctuation
or dispersion around the classical path of a dynamical system. In
presence of strong dissipation these fluctuations are small since it
is well-known (also follows from analysis of
Eqs(\ref{3.13}-\ref{3.17})) that dissipation enhances
classicality\cite{graham}. The role of effective potential of the
similar nature which gives rise to leading order quantum correction
to classical Langevin force had also been noted earlier, e. g., in
the analysis of strong friction limit of quantum stochastic
processes etc. \cite{bao,ank,han11}. Since the corrections are
perturbative in nature they  may differ in form but because of
nonlinearity of the potential they bear close kinship to each other.
We emphasize that the approximate forms of quantum correction must
satisfy the basic symmetry requirement, appropriate equilibrium
distribution and other thermodynamic consistency condition as
pointed out earlier\cite{d1,db4}. In Fig.1 we illustrate the
variation of current as a function of temperature(T) for different
values of the amplitude of external derive($A_0$). One observes that
with increase of $D_q$ (proportional to temperature) the magnitude
of current increases to a maximum followed by decrease and a current
reversal at high temperature. At higher temperature the system is
thermalized as a result of which organized motion in a preferential
direction decreases and the motion towards the load dominates. For
fixed $D_q$ with increase of the value of $A_0$ the magnitude of the
current increases. The effect of quantization of a classical ratchet
is shown in Fig.2, where we present a comparison of the current vs
temperature profile for the classical and the quantum cases. One
observes that at the low temperature region the classical current is
significantly lower in magnitude than quantum current and at the
higher temperature the effect of quantization becomes insignificant.
This may be interpreted in terms of an interplay between quantum
diffusion coefficient $D_q$ and the potential force term $f(x)$.
$f(x)$ contains quantum correction arising due to nonlinearity of
the potential. As temperature $T\rightarrow 0$, $D_q$ approaches to
the value $\frac{1}{2}\hbar\omega_0$, the vacuum limit in deep
tunneling region. The anharmonic terms in $f(x)$ do not contribute
significantly. So the integrand in effective potential $\psi(x)$
increases sharply. On the other hand, as temperature increases,
$D_q$ increases and also $D_q$ and $f(x)$ compete with each other to
merge quantum current to its classical counterpart.

\subsection{Energetics of non-equilibrium fluctuation induced quantum transport}
Efficiency of the ratchet device is an important physical quantity
that quantifies the energetics of non-equilibrium fluctuations in
the transport processes. Depending on the degree and presence of an
external load two distinct approaches have been advocated. It has
been shown that although in many widely accepted cases efficiency is
measured by applying the constant external force, there are
situations, where molecular motors are designed not to pull loads
(e. g., protein transport within a cell). In such cases a minimum
energy input is required to move a particle in a viscous medium. We
therefore discuss the two different situations separately.

\subsubsection{Conventional efficiency in presence of an external load}
 To discuss the energetics of quantum
transport induced by zero mean external derive we consider the
energy transducer which interacts with the external derive and the
load so that the potential takes the following form
\begin{eqnarray}
U(x,t)=V(x)-\int dx \;Q(x
,\langle\delta\hat{x}^n\rangle)+A(t)\;x+lx\nonumber
\\\label{3.22}
\end{eqnarray}
where $V(x)$ is the classical potential, second term represents the
quantum corrections due to nonlinearity of the classical potential
and last two terms are due to external system and the load,
respectively. The interaction of transducer with heat bath is
assumed to be stochastic, as usual. Thus for the movement of
transducer from $x_i(t_i)\rightarrow x_f(t_f)$ the total potential
energy change($\Delta U$) and dissipation energy($E_d$), during the
period $t_i< t <t_f$ are formally given by\cite{sek}

\begin{eqnarray}
\Delta U=U(x_f(t_f),t_f)-U(x_i(t_i),t_i)\label{3.23}
\end{eqnarray}
and
\begin{eqnarray}
E_d=\int_{t_i}^{t_f}[-\dot{x}+\Gamma(t)]dx(t)=\int_{t_i}^{t_f}-\left[
\frac{\partial U(x,t)}{\partial
x}\right]dx(t)\nonumber\\\label{3.24}
\end{eqnarray}
respectively.

Because of the conservation law, the sum of the potential energy
change and dissipation energy must be equal to the total consumption
of energy $E_c$ ($\triangle U+E_d\equiv E_c$), due to the external
system $A(t)$.

\begin{eqnarray}
E_c=\int_{t_i}^{t_f}\frac{\partial U(x,t)}{\partial
t}\;dt\label{3.25}
\end{eqnarray}
In the present case the external system is a periodic function of
time, so that the ensemble average of total consumption
energy($E_c$) and dissipation energy($E_d$) is given by

\begin{eqnarray}
\langle E_c\rangle&=&\int_{t_i}^{t_f}dt \int_{space}\frac{\partial
U(x,t)}{\partial t}P(x,t)dx(t)\nonumber
\\
 &=&\int_{t_i}^{t_f}dt \int _{space}A(t)J(A(t))dx(t) \label{3.26}
\end{eqnarray}
and
\begin{eqnarray}
\langle E_d\rangle=\int_{t_i}^{t_f}dt \int
_{space}dx\left[-\frac{\partial U(x,t)}{\partial
x}\right]J(A(t))\label{3.27}
\end{eqnarray}
respectively. For the square wave with the amplitude $A_0$,
\begin{eqnarray}
\langle (E_c)_{sqr}\rangle=\frac{1}{2} A_0\left[
J(A_0)-J(-A_0)\right]\label{3.28}
\end{eqnarray}

\begin{eqnarray}
\langle (E_d)_{sqr}\rangle&=&\frac{1}{2}
A_0\{J(A_0)-J(-A_0)\}\nonumber
 \\&-&\frac{1}{2}l\{J(A_0)+J(-A_0)
\}\label{3.29}
\end{eqnarray}
Hence the work, that the ratchet system extracts from the external
system $A(t)$ is given by

\begin{eqnarray}
\langle W_{sqr}\rangle&=&\langle (E_c)_{sqr}\rangle-\langle
(E_d)_{sqr}\rangle=\frac{1}{2}\;
l\left[J(A_0)+J(-A_0)\right]\nonumber\\ &=&l\times
J_{sqr}\label{3.30}
\end{eqnarray}
So the work extracted from external system is directly proportional
to square wave current. The conventional efficiency of the ratchet
system is thus calculated on the basis of external load can be
written as

\begin{eqnarray}
\eta &=& \frac{l J_{sqr}}{E_c}\nonumber
\\
 & =&
\frac{l\left[J(A_0)+J(-A_0)\right]}{A_0\left[J(A_0)-J(-A_0)\right]}\label{3.31}
\end{eqnarray}
We now numerically illustrate the behavior of efficiency of the
quantum ratchet system as given above. The effect of quantization of
the reservoir is apparent in Fig.3 in the variation of
efficiency($\eta$) as a function of temperature for different values
of amplitude($A_0$) of the external periodic system. The efficiency
is a decreasing function of temperature for any value of $A_0$ and
it
 decreases with increase of $A_0$ for a fixed value of temperature.
The distinctive behavior of efficiency of a quantum system is
evident from the nature of quantum current that can attain a maximum
value for a finite temperature. On the other hand the maximum
efficiency is realized at the zero strength of thermal fluctuation.
It is thus apparent that the equilibrium fluctuation due to thermal
heat bath is an hindrance for efficient extraction of useful work
from non-equilibrium fluctuations. To have a closer look at the
behavior of efficiency we present in Fig.4 the dissipation energy
and total consumption energy as a function of temperature. At very
low temperature the energy loss due to dissipation($E_d$) during the
movement of energy transducer is small compared to the total energy
consumed from the external system, because for a finite net
displacement $x_i\rightarrow x_f$ it covers minimum path at low
thermal fluctuation. On the other hand at high temperature the path
of energy transducer is more chaotic. So for a finite net
displacement it covers maximum path and loses a greater amount
energy due to dissipation. With increase of the temperature
dissipation energy and total energy consumption from external system
both are increased and the difference between two energies ($E_d$
and $E_c$) become insignificant at higher temperature.

The condition for maximum conventional efficiency can be realized
by rearranging Eq.(\ref{3.31}) as a function of
$\frac{J(-A_0)}{J(A_0)}$
\begin{eqnarray}
\eta=\frac{l}{A_0}\left(1-\frac{2|\frac{J(-A_0)}{J(A_0)}|}{1+|\frac{J(-A_0)}{J(A_0)}|}
\right)\label{3.32}
\end{eqnarray}
In the limit $|\frac{J(-A_0)}{J(A_0)}|\rightarrow 0$, the maximum
efficiency of the energy transform for a given load and force
amplitude is given by (the limit can be achieved by suitable
adjustment of parameters)

\begin{eqnarray}
\eta_{max}=\frac{l}{A_0}\label{3.33}
\end{eqnarray}
Now we have two important conclusions regarding the maximum
efficiency of a ratchet system, (i) it is a simple ratio of load to
a parameter of external system (strength of external system) and it
is independent of the characteristics of the bath. (ii) $\eta_{max}$
being independent of the nature of the bath and the system
potential, is same both in quantum and classical systems.

 In Fig.5 we compare the conventional efficiency vs temperature profile for
the classical and the quantum cases for different values of $A_0$.
We observe that the efficiency of quantum ratchet is significantly
lower than classical one and the difference becomes insignificant at
higher temperature. Since the vacuum fluctuations tend to be
effective in the quantum system as one approaches the zero
temperature limit, the transducer loses a higher amount of
dissipation energy than the classical one.

\subsubsection{Efficiency in absence of an external load and generalized efficiency}
We now consider the situation where the motor works without any
external load. The task is not only to translocate the motor over
a distance $L$ but also to do this with a given average velocity
it must work against the viscous force $\gamma \langle v
\rangle$. Now replacing the load by $\gamma \langle v \rangle$ we
can define an efficiency (Stokes efficiency)

\begin{eqnarray}
\eta_S &=& \frac{\gamma \langle v \rangle ^2}{E_c}\label{3.34}
\end{eqnarray}
By combining the contribution due to (\ref{3.31}) and
(\ref{3.34}), it is possible to define further a generalized
efficiency for the quantum system
\begin{eqnarray}
\eta_G &=& \frac{lJ_{sqr}+\gamma \langle v \rangle
^2}{E_c}\label{3.35}
\end{eqnarray}
The above expression is the quantum generalization of the
classical generalized efficiency as given earlier by Suzuki and
Munakata \cite{dai} and Der\'{e}nyi \textit{et al} \cite{der}.
This account for both the work that the motor performs against the
external load $l$ as well as the work that is necessary to move
the particle over a given distance in a viscous environment at
the average velocity $\langle v \rangle$.

In Fig.6 we the present variation of Stokes efficiency $\eta_S$ as
a function of temperature. It is important to observe  that
efficiency reaches a maximum at a particular temperature both for
classical as well as for the quantum case. However, again at low
temperature the efficiency of the classical system drops to zero
in sharp contrast to quantum case. At high temperature the system
however, tends to classical regime as expected.

\section{conclusion}
In this paper we have calculated the efficiency of a forced thermal
ratchet in a quantum mechanical context. We have shown that the
quantum current is markedly higher as compared to classical current
at low temperature while the difference becomes insignificant at
higher temperature. In contrast to the behavior of quantum current
at low temperature, the conventional efficiency of a classical
ratchet in presence of a load is higher at low temperature as
compared to its quantum counterpart and again the efficiency in the
two cases tends to merge at higher temperature. Furthermore the
maximum efficiency is independent of the nature of the system
potential and the bath and is thus independent of quantization. We
have also examined a quantum version of Stokes efficiency in absence
of load where energy due to frictional resistance is considered as a
part of expenditure of useful energy. A significant quantum
enhancement of Stokes efficiency at low temperature has been
observed. The careful consideration of the total energy consumption
and dissipation reveals that the generation of higher current and
Stokes efficiency may not always imply the higher efficiency of
thermal ratchet in a conventional sense although the generic
features of the device in its classical and quantum versions remain
the same. \acknowledgments Thanks are due to the Council of
Scientific and Industrial Research, Govt. of India, for a fellowship
(PKG).
\appendix
\begin{appendix}
\section{The passage from Eq.(2.13) to Eq. (2.14)}
We start from basic definition\cite{loui}
\begin{equation}\label{1}
2D_q=\frac{1}{2\Delta{t}}\int_t^{t+\Delta{t}}dt_1\int_t^{t+\Delta{t}}dt_2\;\;
\langle\Gamma(t_1)\Gamma(t_2)\rangle_s
\end{equation}
Using Eq.(2.13)  in Eq.(1) yields
\begin{eqnarray}\label{2}
2D_q&=&\frac{1}{2\Delta{t}}\int_0^\infty
d\omega\;\kappa(\omega)\;\rho(\omega)\hbar\omega\;\coth\left(\frac{\hbar\omega}{2kT}\right)
\nonumber
\\
&\times&\int_t^{t+\Delta{t}}dt_1\int_t^{t+\Delta{t}}dt_2\;\cos\omega(t_1-t_2)
\end{eqnarray}
Explicit integration over time gives
\begin{equation}\label{3}
2D_q=\frac{1}{2\Delta{t}}\int_0^\infty
d\omega\;\kappa(\omega)\;\rho(\omega)\hbar\omega\;\coth\left(\frac{\hbar\omega}{2kT}
\right)\;\;\;I(\omega,\Delta{t} )
\end{equation}
where
\begin{equation}\label{4}
I(\omega,\;\Delta{t})=\frac{4}{w^2}\;\;\sin^2\frac{\omega\Delta{t}}{2}
\end{equation}
Putting $\kappa(\omega)\;\rho(\omega)=\frac{2}{\pi}\;\gamma$, we
obtain
\begin{equation}\label{5}
 2D_q=\frac{\gamma}{\Delta{t}\;\pi}\int_0^\infty
d\omega\;\hbar\omega\;\coth\left(\frac{\hbar\omega}{2kT}\right)\;\;
\frac{\sin^2\frac{\omega\Delta{t}}{2}}{(\frac{\omega}{2})^2}
\end{equation}
Following Louisell\cite{loui} we have under Markovian condition, the
correlation time $\tau_c\;\ll\;\Delta{t}$, the coarse-grain
time(over which the probability distribution function evolves). Thus
as $\Delta{t}\rightarrow \infty$(in scale of $\tau_c$ which goes to
zero) the function
$\frac{\sin^2\frac{\omega\Delta{t}}{2}}{(\frac{\omega}{2})^2}$
oscillates violently so that one takes the slowly varying quantity
$[\hbar\omega\;\coth\frac{\hbar\omega}{2kT}]$ out of the integration
over frequency with an average value
$\hbar\omega_0\coth\frac{\hbar\omega_0}{2kT}$, $\omega_0$ be an
average static frequency. Since the integral
$\int_\infty^\infty\frac{\sin^2x\Delta{t}}{x^2}dx=\pi\Delta{t}$ it
follows immediately from Eq.(5)
\begin{equation}\label{6}
2 D_q=\gamma\; \hbar \omega_0 \coth\left(\frac{\hbar \omega_0}{2 k
T}\right)
\end{equation}
as given in Eq.[2.15]. Again starting from Eq.(2.13), we use the
same argument as before to have
\begin{eqnarray}\nonumber
\langle \Gamma(t_1)\Gamma(t_2) \rangle_s &=&
\frac{1}{2}\int_0^\infty d\omega\; \kappa(\omega)\rho(\omega) \hbar
\omega\nonumber \\
 &\times&\coth\left(\frac{\hbar \omega}{2 k
T}\right) \cos\omega(t_1-t_2)
\end{eqnarray}
and we use
\begin{eqnarray}
\int_0^\infty d\omega \cos \omega \tau = \pi\; \delta(\tau)\nonumber
\end{eqnarray}
to obtain
\begin{eqnarray}
\langle \Gamma(t_1)\Gamma(t_2) \rangle_s &=& \frac{1}{2}
\int_0^\infty d\omega \left[\frac{2}{\pi} \gamma \right] \hbar
\omega\nonumber
\\
&\times&\coth\left(\frac{\hbar \omega}{2 k
T}\right) \cos\omega(t_1-t_2)\nonumber\\
&=& \frac{\gamma\;\hbar \omega_0}{\pi}\coth\left(\frac{\hbar
\omega_0}{2 k T}\right) \pi \delta(t_1-t_2)\nonumber\\
&=&\gamma\; \hbar \omega_0 \coth\left(\frac{\hbar \omega_0}{2 k
T}\right) \delta(t_1-t_2)\label{7}
\end{eqnarray}
Therefore from Eq.(6) and Eq.(7) we have
\begin{eqnarray}
\langle \Gamma(t_1)\Gamma(t_2) \rangle_s &=& 2D_q
\delta(t_1-t_2)\nonumber
\end{eqnarray}
Thus the derivation within Markovian approximation clearly depends
on the time scale separation. The results are valid even at absolute
zero as emphasized by Louisell.
\end{appendix}

\begin{center}
{\bf Figure Captions}
\end{center}
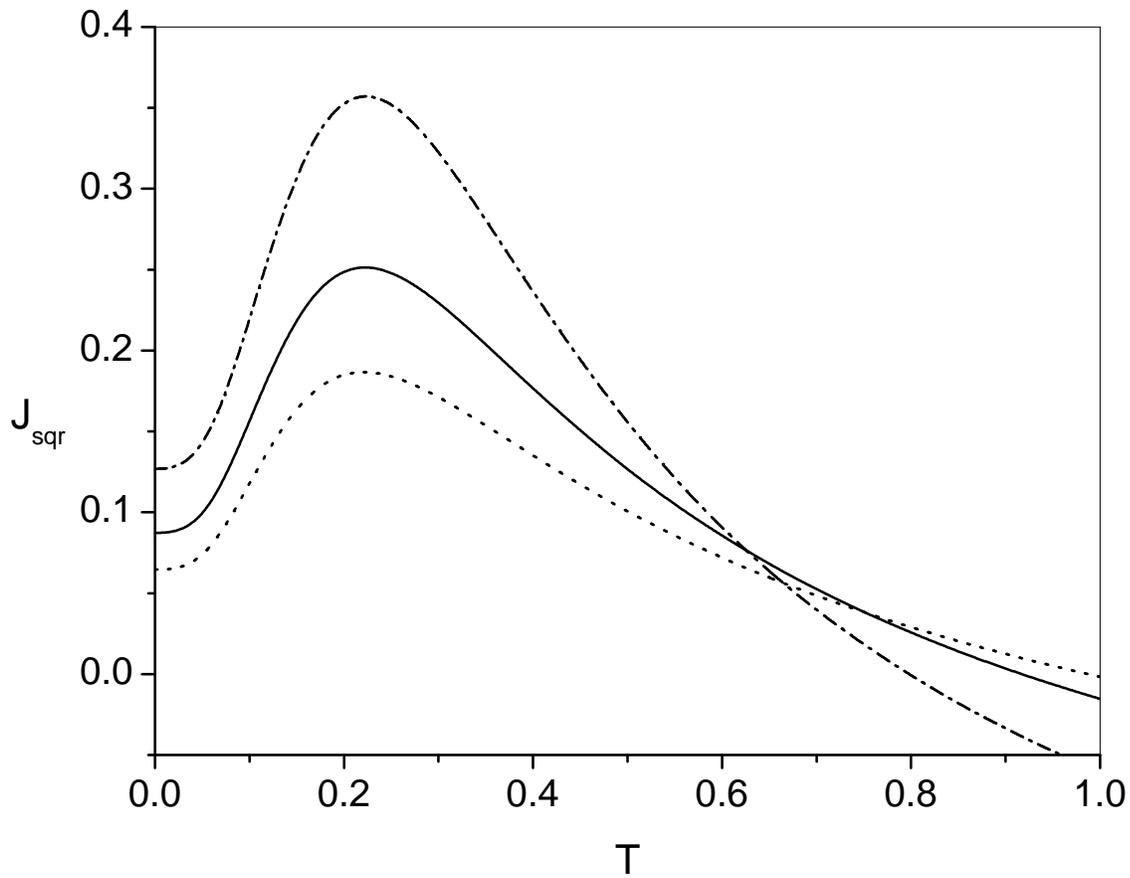
Fig.1. A plot of quantum current vs $T$ for different strength of
external periodic force (i) $A_0=1.0$ (dotted line), (ii) $A_0=1.2$
(solid line), (iii) $A_0=1.5$ (dash-dot line) and $l=0.01$.(All the
quantities are dimensionless)

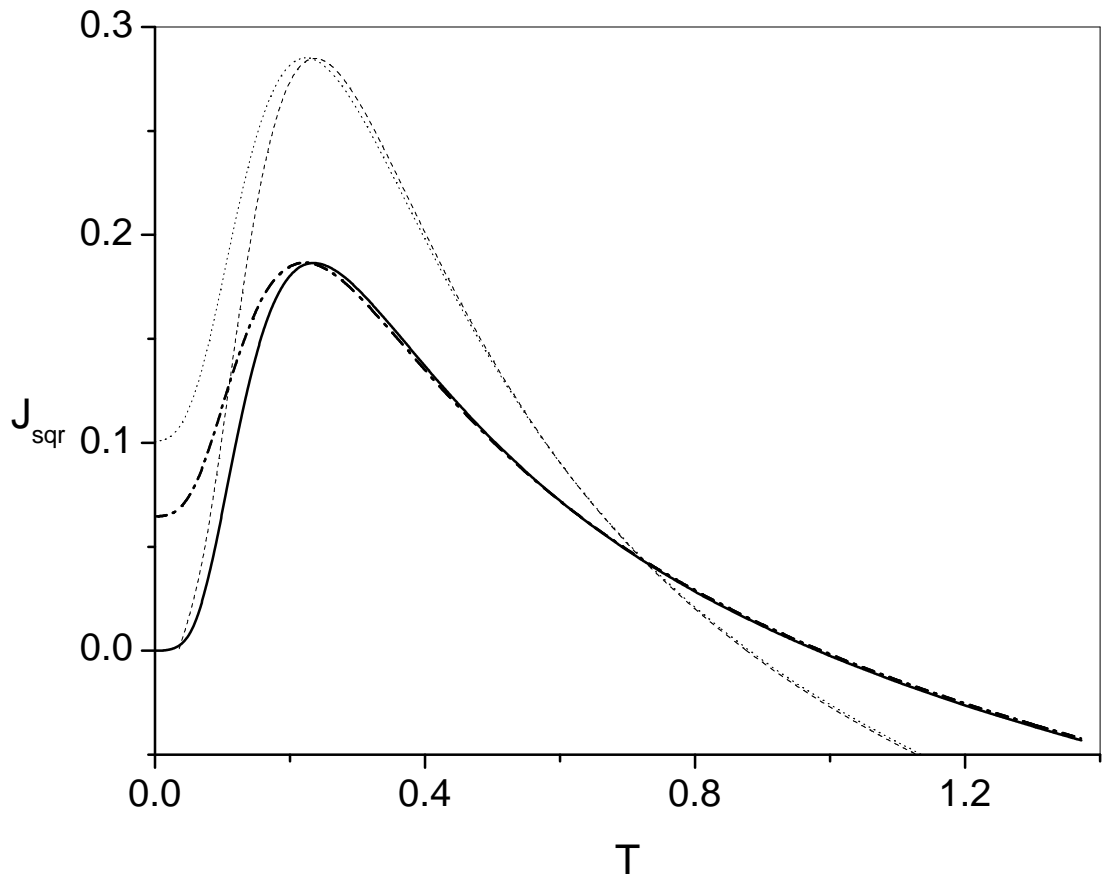
Fig.2. A current$J_{sqr}$ vs temperature($T$) plot comparing
classical(dashed line and  solid line) and quantum (dotted line and
dashed dot line) limit for different strength of external periodic
force (i) $A_0=1.3$ (dotted line and dashed line), (ii) $A_0=1.0$
(dashed dot line and solid line) and $l=0.0$.(All the quantities are
dimensionless)

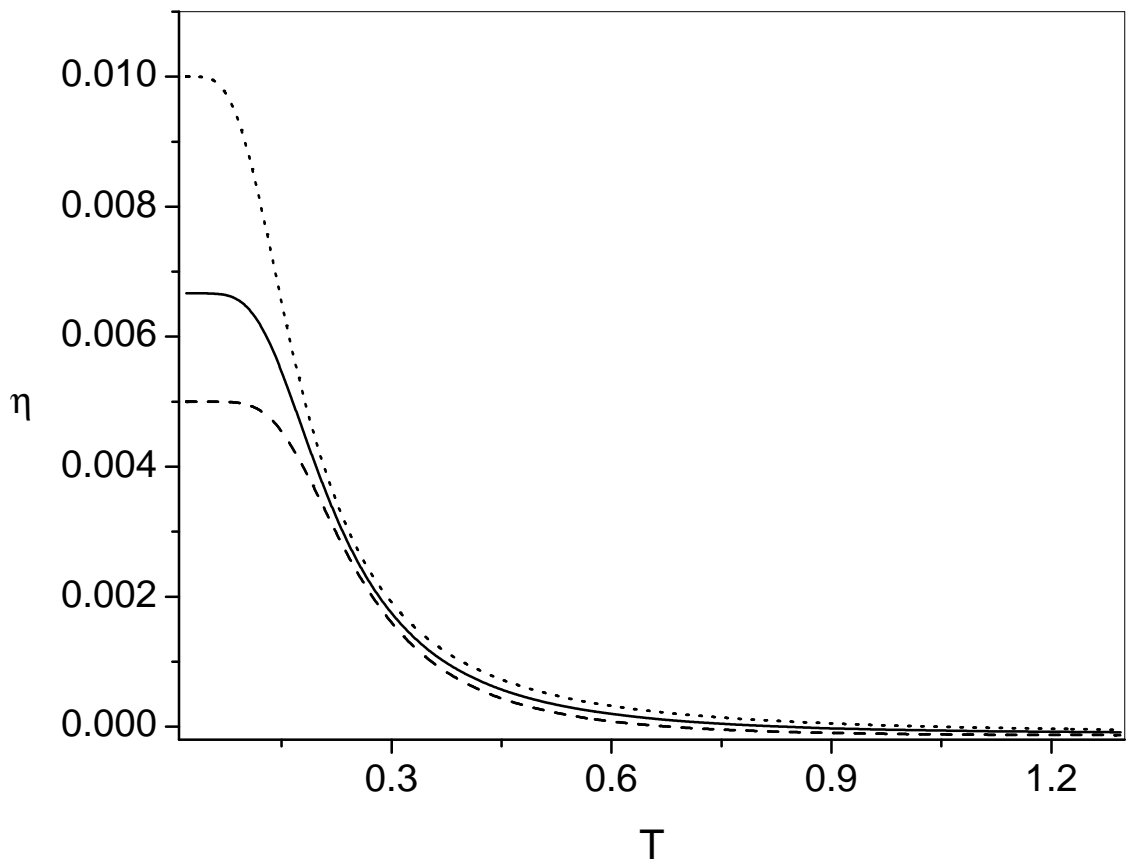
Fig.3. A plot of quantum efficiency vs $T$ for different strength of
external periodic force (i) $A_0=1.0$ (dotted line), (ii) $A_0=1.5$
(solid line), (iii) $A_0=2.0$ (dashed line) and $l=0.01$.(All the
quantities are dimensionless)

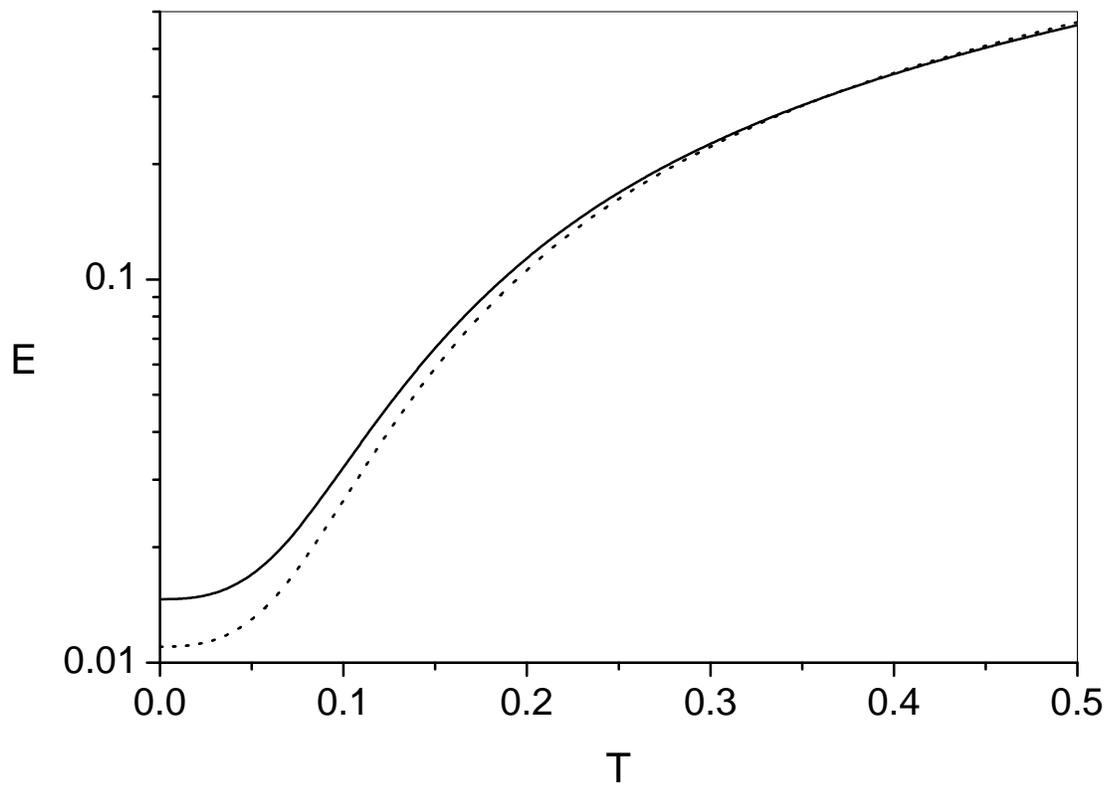
Fig.4. A comparison between dissipation energy $E_d$(dotted line)
and total energy consumption $E_c$ (solid line) as a function of
temperature for the parameter set $l=0.05$ and $A_0=1.0$.(All the
quantities are dimensionless)

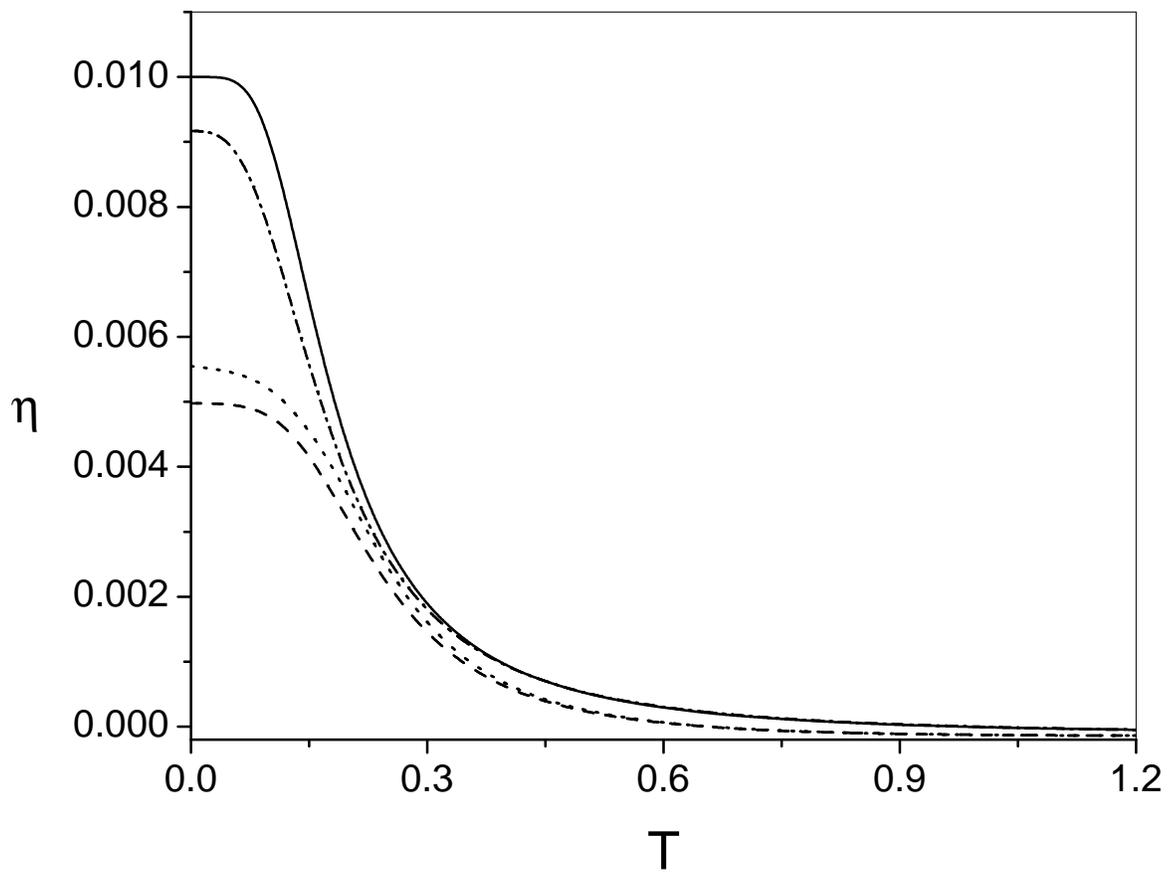
Fig.5. Conventional efficiency($\eta$) vs temperature($T$) plot
comparing classical(dotted line and solid line) and quantum (dashed
line and dashed dot line) limit for different strength of external
periodic force (i) $A_0=1.3$ ( dotted line and dashed line ), (ii)
$A_0=1.0$ (solid line and dashed dot line) and $l=0.0$.(All the
quantities are dimensionless).

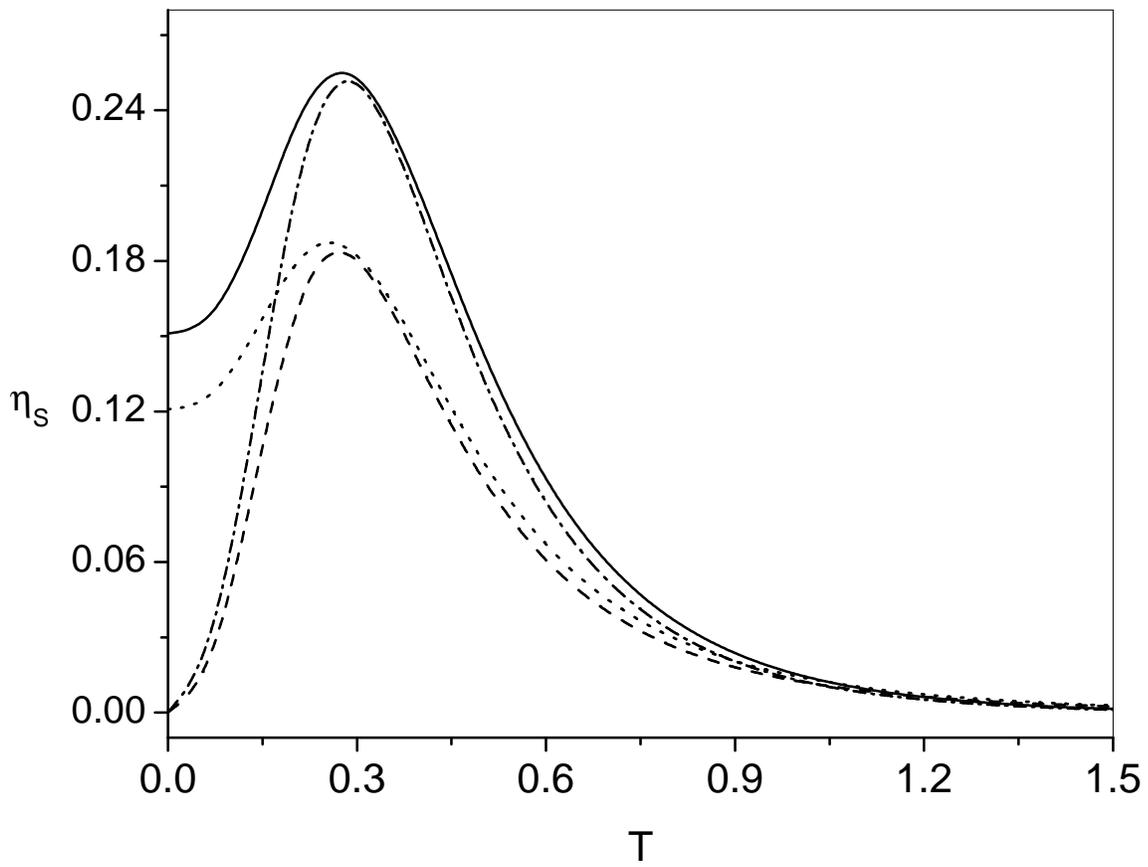
Fig.6. A comparison between classical(dashed dot line and dashed
line) and quantum (solid and dotted line) Stokes efficiency for
different strength of external periodic force (i) $A_0=1.3$ (solid
line and dashed dot line), (ii) $A_0=1.0$ (dotted line and dashed
line) and $l=0.0$.(All the quantities are dimensionless)

\end{document}